\documentclass[11pt]{article}
\usepackage{amsfonts}
\usepackage{amssymb}
\usepackage{amsmath}
\usepackage{epsfig}
\usepackage{graphicx}
\usepackage{bm}
\usepackage{color}
\usepackage{hyperref}

\newlength{\bredde}
\def\slash#1{\settowidth{\bredde}{$#1$}\ifmmode\,\raisebox{.15ex}{/}
\hspace*{-\bredde} #1\else$\,\raisebox{.15ex}{/}\hspace*{-\bredde} #1$\fi}
\newcommand{\be}{\begin{equation}}
\newcommand{\ee}{\end{equation}}
\newcommand{\bea}{\begin{eqnarray}}
\newcommand{\eea}{\end{eqnarray}}

\newcommand{\eins}{\leavevmode\hbox{\small1\kern-3.8pt\normalsize1}}

\date{}

\begin{document}
\topmargin -1.4cm
\oddsidemargin -0.8cm
\evensidemargin -0.8cm
\title{\Large{{\bf
 Thermal transport through non-ideal Andreev quantum dots
}}}

\vspace{1.5cm}

\author{~\\{\sc Pedro Vidal}%$^{1}$
\\~\\
%$^1$
Fakult\"at f\"ur Physik,
Universit\"at Bielefeld,\\
D-33501 Bielefeld, Germany\\~\\
}

\maketitle
\vfill
\begin{abstract}
We consider the scenario of thermal transport through two types of Andreev quantum dots
which are coupled to two leads, belonging to the  
Class D and Class C symmetry classes.
Using the random matrix description 
we derive the joint probability density function (j.p.d.f.) in term of Hypergeometric Function
of Matrix Arguments when we consider one lead to be attached ideally and one lead non 
ideally. For the class C ensemble we derive a more explicit
representation of the j.p.d.f.
which results in a new type of random matrix model.\\
%\vskip 0.1cm\ \\
%{\bf PACS:} 02.10.Yn, 02.30.Cj, 02.50.Sk\\
%{\bf MSC:} 15B52, 33E20, 60B20  
\end{abstract}
\vfill

\thispagestyle{empty}
\newpage

\renewcommand{\thefootnote}{\arabic{footnote}}
\setcounter{footnote}{0}

%%%%%%%%%%%%%%%%%%%%%%%%%%%%%%%%%%%%%%%%%%%%%%%%%%%%%%%%%%%%%%%%%%%%%%%%%%%
\section{Introduction}

Quantum dots form an important class of mesoscopic systems 
whose electric and thermal transport properties 
are being actively studied. 
Random matrix theory has had a tremendous success in describing 
these systems in the limit of low voltage and low temperature, when the 
classical motion of an electron is chaotic in the dot.
In this regime transport through the dot 
boils down to characterizing how  single (quasi-)particles
are transmitted through a chaotic cavity. 
The Landauer-B\"uttiker approach does this trough 
the study of the scattering matrix. 
If the classical dynamics 
is chaotic in the cavity then the scattering matrix
would be well described by a random uniformly distributed unitary matrix \cite{BlumelSmilansky}, an ensemble which 
had already been studied by Dyson.
Taking the distribution to be uniform over the scattering matrix 
 gave rise to three ensembles,
 namely the circular unitary, orthogonal and symplectic
 ensemble or CUE, COE and CSE.
Each of these is determined by the presence or absence of time reversal symmetry 
and spin rotation symmetry of the electron in the chaotic cavity.
In \cite{AltlandZirnbauer} Altland and Zirnbauer showed the existence of four more types of symmetry classes
which appear when particle-hole symmetry is present.
When a quantum dot is put in contact with a superconductor
an electron moving inside the cavity can be reflected as a hole.
This process is called Andreev reflection \cite{Andreev} and in \cite{AltlandZirnbauer}
it was shown that these hybrid normal metal-superconductor systems
formed four new symmetry classes named class D, C, DIII and CI.
When considering a quantum dot within these symmetry classes and assuming that the scattering matrix is uniformly
distributed
these give rise to the Circular Real Ensemble (CRE)
, Circular Quaternion Ensemble (CQE), 
Circular Real Time reversal symmetric Ensemble  (T-CRE)
and the 
Circular Quaternion Time reversal symmetric Ensemble (T-CQE), \cite{Serban,dahlhaus}. 
Since Andreev reflections change the charge of the
 particle moving in the cavity, electric transport 
is no longer the same as thermal transport, which in these systems is the same as particle transport. 
Put differently, a particle scattered through the cavity will transport a definite amount
of energy  but not a definite amount of charge since it
can come out as an electron or a hole.

The uniform distribution over all of these circular ensembles is
the most ``simple'' scenario which, although it 
can be realized experimentally, need not be the case.
It was shown in \cite{MelloPereyraSeligman} that, if on average 
the scattering matrix was different from zero then the
distribution over the scattering matrix is given by the Poisson Kernel, $P(S)$. 
This was shown for the CUE ,COE and CSE cases.
A generalization of the Poisson Kernel for the CRE , CQE ,T-CRE and T-CQE was derived
 in \cite{Beri}.
When the distribution is given by  the Poisson kernel, or its generalization,
the system is said to be non ideal.
When the leads are ideally coupled to the chaotic cavity
the scattering matrix distribution is uniform.

Aside from being a more general description of the quantum dots,
non ideal systems can be attractive for different reasons.
For example, in \cite{Vivo} it was shown that semi-non-ideal quantum dots could be used to 
tune the amount of entanglement between two electrons scattering on the quantum dot.

Many transport observables, such as the conduction or the shot noise,
can be written down in terms of the transmission or the reflection eigenvalues.
The main obstacle when studying non ideal scenarios is 
that the joint probability density function (j.p.d.f.) 
for these eigenvalues is not available while it is available for the ideal case.
Exceptions to this are the cases of the semi-non-ideal quantum dot
with 
broken/preserved time reversal symmetry and spin rotation symmetry.
These are non ideal versions of the CUE, solved in \cite{VidalKanzieper} and studied in \cite{Vivo} 
regarding entanglement and the COE and CSE analyzed in \cite{JaroszVidalKanzieper}.

We will consider the problem of thermal transport through 
Andreev quantum dots where two leads are attached 
and we will consider only one lead to be non ideal (semi-non-ideal quantum dot). 
The symmetry classes we will analyze are the 
Class D and C.
Class D systems correspond to those with broken time-reversal and spin-rotation symmetry while class 
C has only broken time-reversal symmetry. 
This means we are looking at the non ideal version of the CRE and CQE,
and refer to them as the Poisson Real and Quaternion Ensemble (PRE and PQE).
In these cases the scattering matrices are 
orthogonal ($O(N)$) and symplectic ($Sp(N)$)respectively.
 The orthogonal scattering matrices can be further split into 
two parts. Matrices with determinant $1$ and matrices with determinant $-1$.
The determinant is called the topological quantum number
and when it is $1$ ($-1$) we are in the topologically (non-)trivial phase. 
We will consider the case when the determinant is $1$ and thus
the scattering matrices form the group $SO(N)$.

Our strategy is analogous to the one used in previous work \cite{JaroszVidalKanzieper}.
In section \ref{sec:LB} we will review the Landauer-B\"uttiker approach
and explain where the main hurdle lies to find the j.p.d.f.. 
In section \ref{sec:jpdf} we will use the theory of symmetric function
to derive how the j.p.d.f. can be expressed in terms of Hypergeometric Function of Matrix Arguments (HFMA).
In \ref{sec:TSF} we review the main results  from the theory of symmetric functions
that we use in this derivation.
In section \ref{sec:representationHFMA} we will derive a representation
of the HFMA which  will be useful to derive a more compact representation to the
j.p.d.f. for the quaternion ensemble.

\section{Landauer-B\"uttiker approach}
\label{sec:LB}
The system we consider is an Andreev quantum dots 
with a left lead with $n$ channels and a right lead 
with $m$ channels. We take $n\leq m$ and for the real ensemble we have included the spin  and particle/hole quantum numbers.
For the quaternion ensemble we do not include spin quantum number.
The scattering matrix, $S$, is then a $(n+m)\times (n+m)$ matrix for the real ensemble
and for the quaternion ensemble it is a $(n+m)\times (n+m)$ 
matrix with quaternion elements.
Which means the scattering matrix is either an orthogonal matrix in $SO(N)$ or a symplectic matrix in $Sp(N)$. 
The transmission matrix ,${\bf t}_{n\times m}$,  is a sub-block of the scattering matrix. 
\begin{align}
 S=\left(\begin{array}{cc}
    {\bf r}_{n\times n} & {\bf t}_{n\times m} \\
    {\bf t'}_{m\times n}& {\bf r'}_{m\times m} 
   \end{array}\right) .\label{eq:S}
\end{align}
The Landauer-B\"uttiker approach characterizes transport
through a quantum dot by the eigenvalues 
of the product of the transmission matrix with its hermitian conjugate.
That is to say, the eigenvalues $T_{j}$ of the matrix $\bf{tt^{\dagger}}$ 
determine 
the thermal transport observables, such as the
conductance $G$, through the following formula:
\begin{align}
G=dG_{0}\sum_{j=1}^{n}T_{j} ,
\end{align}
with $G_{0}=\frac{\pi^{2}k_{B}^{2}T_{0}}{6h}$ and $d$ denotes the degeneracy of the transmission eigenvalues.
Alternatively the reflection eigenvalues $R_{j}$, the eigenvalues of the matrix $\bf{rr^{\dagger}}$, 
 can be used. 
 They are related to the transmission eigenvalues as $R_{j}=1-T_{j}$ and we will use these instead of the 
 transmission eigenvalues.
The Random Matrix Theory description of quantum dots start with a given distribution, $P(S)$, over
the scattering matrix.
Given this  distribution 
the expectation value of an observable depending on the transmission eigenvalues,
 $F(R_{j})$,
 is given by
\begin{align}
\left<F(R_{j})\right>=\int d\mu(S) P(S) F(R_{j}),
\end{align}
where $d\mu(S)$ is the uniform measure or Haar measure over $S$. In order to characterize the statistics
of observables depending on the reflection eigenvalues one needs to derive from 
$P(S)$ the joint probability density function (j.p.d.f.) of the reflection eigenvalues,
$\mathcal{P}(R_{j})$.

\noindent By using the polar decomposition of the scattering matrix it is parametrized as follows
\begin{align}
 S&=\left(\begin{array}{cc}
    U_{1} & 0\\
    0& U_{2} 
   \end{array}\right) 
   \left(\begin{array}{cc}
     r & t \\
    -t'& r' 
   \end{array}\right)
   \left(\begin{array}{cc}
    V_{1}^{\dagger} & 0 \\
    0& V_{2}^{\dagger} 
   \end{array}\right)\label{eq:polardecomposition}\\
   &=U 
   \left(\begin{array}{cc}
     r & t \\
    -t'& r' 
   \end{array}\right)
V^{\dagger}\notag
\end{align}
where $r,t,t'$ and $r'$ are diagonal matrices. 
$r$ has as diagonal values $r_{1},\cdots r_{n}$, with $r_{j}\in \left[0,1\right]$, and 
the reflection eigenvalues $R_{j}$ are given by $r_{j}^{2}$.
$r'$ the same values as $r$ and an extra $m-n$ of $1$'s as diagonal values.   
 \begin{align}
 r&=\text{diag}\left\{r_{1},\cdots , r_{n}\right\}\notag\\
 r'&=\text{diag}\left\{r_{1},\cdots , r_{n},1\cdots 1\right\}\notag
\end{align}
$t$ on the other hand is rectangular ($n\times m$) with $n\leq m$ and has as diagonal elements $t_{1},\cdots t_{n}$.
While $t'$ is the transpose of $t$ , $t'=t^{T}$.
In order for this parametrization to be unique we take $U_{1},V_{1}\in O(n)$ , $V_{2}\in O(m)$ and $U_{2}\in O(m)/O(m-n)$ when $S\in SO(N)$.
When $S\in Sp(N)$ we take $U_{1}\in Sp(n)/Sp(1)^{n}$, $V_{1}\in Sp(n)$ , $V_{2}\in Sp(m)$ and $U_{2}\in Sp(m)/Sp(m-n)$.
However, as shown in section \ref{sec:polardecomposition},
the integrals over the coset spaces can be extended to the full group once the Jacobian is computed.
Additionally, for the orthogonal matrices the determinant is  equal to $1$ 
 and we need to insure 
that this condition is fulfilled
in the parametrization.
The determinant is given by 
\begin{align}
\det\left[
 \left(\begin{array}{cc}
    U_{1} & 0\\
    0& U_{2} 
   \end{array}\right) 
   \left(\begin{array}{cc}
     r & t \\
    -t'& r' 
   \end{array}\right)
   \left(\begin{array}{cc}
    V_{1}^{\dagger} & 0 \\
    0& V_{2}^{\dagger} 
   \end{array}\right)\right]
 &   = \det\left[U_{1}\right]
  \det\left[U_{2}\right]
 \det\left[V_{1}\right]
 \det\left[V_{2}\right].
\notag
\end{align}
Therefore we need to insure
\begin{align}
 \det\left[U_{1}\right]
  \det\left[U_{2}\right]
 \det\left[V_{1}\right]
 \det\left[V_{2}\right]=+1 .\notag
\end{align}
We set in our parametrization 
$\det\left[V_{2}\right]$ equal to $\det\left[V_{1}\right]$
and $\det\left[U_{2}\right]$ equal to $\det\left[U_{1}\right]$.
Meaning the matrices $U_{2}$ and $V_{2}$ are constricted.
Denoting by $J(r_{j})$ the Jacobian of the transformation of Eq. (\ref{eq:polardecomposition}) 
one gathers
\begin{align}
\left<F(r_{j})\right>=&
\prod_{j=1}^{n}\int_{0}^{1} dr_{j}
F(r_{j})
J(r_{j})\int d\mu\left(U\right)d\mu\left(V\right)
P\left(U 
   \left(\begin{array}{cc}
     r & t \\
    -t'& r' 
   \end{array}\right)
V^{\dagger}\right)
\label{eq:expectationvlue}
\end{align}
For the quaternion ensemble $d\mu\left(U\right)d\mu\left(V\right)$ is a product of Haar measures over independent matrices
\begin{align}
 d\mu\left(U\right)d\mu\left(V\right)
 = d\mu\left(U_{1}\right)d\mu\left(U_{2}\right)
 d\mu\left(V_{1}\right) d\mu\left(V_{2}\right) \label{eq:measureq}
\end{align}
For the orthogonal  ensemble $d\mu\left(U\right)d\mu\left(V\right)$ is a product of Haar measures over matrices
whose determinants are related
\begin{align}
 d\mu\left(U\right)d\mu\left(V\right)
 = d\mu\left(U_{1}\right)d\mu\left(U_{2}\right)
 d\mu\left(V_{1}\right) d\mu\left(V_{2}\right)
 \delta\left(\det\left[V_{1}\right]-\det\left[V_{2}\right]\right)
 \delta\left(\det\left[U_{1}\right]-\det\left[U_{2}\right]\right) \label{eq:measureo}
\end{align}
with  $U_{j},V_{j}\in O(n)$ 
for the real ensemble
and $U_{j},V_{j}\in Sp(n)$ for the quaternion ensemble.
We note also that 
for the quaternion ensemble the matrix is made of quaternion elements and so the diagonal 
quaternion matrix $r$ of singular values has $n$ blocks $r_{j}\mathbb{I}_{2\times 2}$.
Thus the singular values $r_{j}$ are double degenerate.
The j.p.d.f., denoted by $ \mathcal{P}(R_{j})$, can almost be read of Eq. (\ref{eq:expectationvlue}).
Given that $r_{j}^{2}=R_{j}$ we only need to make a change of variables in Eq. (\ref{eq:expectationvlue}).
\begin{align}
\left<F(R_{j})\right>=&
\left<F(r^{2}_{j})\right>=
\prod_{j=1}^{n}\int_{0}^{1} dr_{j}
F(r^{2}_{j})
J(r_{j})\int d\mu\left(U\right)d\mu\left(V\right)
P\left(U 
   \left(\begin{array}{cc}
     r & t \\
    -t'& r' 
   \end{array}\right)
V^{\dagger}\right)
\notag\\
=&
\prod_{j=1}^{n}\int_{0}^{1} dR_{j}
F(R_{j})
J\left(\sqrt{R_{j}}\right)
(2R_{j})^{-\frac{1}{2}}
\int d\mu\left(U\right)d\mu\left(V\right)
P\left(U 
   \left(\begin{array}{cc}
     \sqrt{R} & \sqrt{T} \\
    -\sqrt{T'}& \sqrt{R} 
   \end{array}\right)
V^{\dagger}\right)
\end{align}
The j.p.d.f. is then given by
\begin{align}
 \mathcal{P}(R_{j})=&
 J\left(\sqrt{R_{j}}\right)
(2R_{j})^{-\frac{1}{2}}
\int d\mu\left(U\right)d\mu\left(V\right)
P\left(U 
   \left(\begin{array}{cc}
     \sqrt{R} & \sqrt{T} \\
    -\sqrt{T'}& \sqrt{R} 
   \end{array}\right)
V^{\dagger}\right).
\label{eq:integraljpdf}
\end{align}
When both leads are attached to the quantum dot ideally 
random matrix theory models the ensemble of scattering
through circular ensembles, meaning $P(S)=1$.
The ensemble generated by the orthogonal matrices is then called the Circular Real Ensemble (CRE)
and  the one generated by the symplectic matrices is called the Circular Quaternion Ensemble (CQE).
The j.p.d.f. for this case was derived in \cite{ForresterQuantum}, \cite{dahlhaus}.  
\begin{align}
\mathcal{P}(R_{j})\propto &|\Delta(R_{j})|^{\beta} \left(1-R_{j}\right)^{\frac{\beta}{2}\left(m-n+1\right)-1}R_{j}^{\frac{\eta}{2}}\notag
\end{align}
with the following values of $\beta,\eta$ and $d$ depending on the ensemble 
\begin{align}
 \begin{array}{c|c|c|c}
    \text{Ensemble}  & \beta & \eta & d\\
  \text{CRE} & 1 & -1& 1\\
  \text{CQE} & 4& 2 & 4\\
 \end{array}\notag
\end{align}
A more general situation is described when one allows for a non-ideal coupling 
between the leads and the dot.
In this situation the distribution over the scattering matrix 
is a Poisson type kernel \cite{Beri}
\begin{align}
 P(S)=&\frac{1}{\mathcal{C}(\hat{\gamma})}|\det\left[1-\hat{\gamma} S\right]|^{-N_{\sigma}}, \label{eq:poissonkernel}\\
 N_{\sigma}=&N+\sigma \notag
\end{align}
where $N=n+m$ for the real ensemble and $N=2n+2m$ for the quaternion ensemble.
$\mathcal{C}(\hat{\gamma})$ is the normalization constant to be computed later on.
These ensembles are no longer circular and we will refer to them as the Poisson Real Ensemble (PRE)
and the Poisson Quaternion Ensemble (PQE).
The parameter $\sigma$  depends on the ensemble is given below.
\begin{align}
\begin{array}{cc}
 \text{PRE} & \sigma=-1\\
 \text{PQE} & \sigma=1
\end{array}\label{eq:list} 
\end{align}
The matrix $\hat{\gamma}$ encodes the coupling 
between the left/right lead and the dot.
\begin{align}
  \hat{\gamma}=&
 \left(
 \begin{array}{cc}
  \hat{\gamma}_{L}& 0\\
0& \hat{\gamma}_{R}
\end{array}
\right)\notag
\end{align}
The left lead is taken to be non ideally coupled, $\hat{\gamma}_{L}\neq 0$,
while the right one is arbitrary $\hat{\gamma}_{R}= 0$. 
We call this the semi-non-ideal scenario.
 Since we are studying the case 
where the right lead has coupling $\hat{\gamma}_{R}=0$
the Poisson like kernel 
simplifies to
\begin{align}
 P(S)=&
 \frac{1}{\mathcal{C}(\hat{\gamma})}
 |\det\left[1-\hat{\gamma}_{L} {\bf r}\right]|^{-N_{\sigma}} \notag
\end{align}
For the semi-non-ideal system we gather then from Eqs. (\ref{eq:integraljpdf}), (\ref{eq:measureq}) and (\ref{eq:measureo})
\begin{align}
\mathcal{P}\left(R_{j}\right)
 =&
  J\left(\sqrt{R_{j}}\right)
(2R_{j})^{-\frac{1}{2}}
 \int_{} d\mu(U_{1})d\mu(V_{1})
 \left|
 \det\left[1-\gamma_{L} U_{1}rV_{1}^{\dagger}\right]\right|^{-(N+\sigma)}, \label{eq:jpdf} 
\end{align}
where the integrals are either over $O(n)$ or $Sp(n)$.
The problem of finding the j.p.d.f. thus boils
down to performing the integration over the orthogonal/symplectic group.
Since we rely heavily on the theory of symmetric functions we 
have included an appendix where the most important 
features of the theory, for our present calculations,
are explained.

\section{The joint probability density function}
\label{sec:jpdf}
\subsection{Poisson Real Ensemble }
For both ensembles the strategy is the same but we will perform them 
separately for the sake of clearness. The idea is to expand the Poisson kernel in terms of symmetric functions
in order to perform the integrations over the group. 
Once this is done the result will turn out to be known as Hypergeometric Functions of Matrix Argument(s) (HFMA).
In section \ref{sec:representationHFMA} we will elaborate on different representations of these HFMA.

\noindent For the PRE we expand the inverse determinant using Eq. (\ref{eq:detexpansion}) in terms 
of the schur functions $S_{\lambda}\left(X\right)$. 
The integral to be performed is denoted by $I_{PRE}(\hat{\gamma},R_{j})$ and defined as follows:
\begin{align}
 I_{PRE}\left(\hat{\gamma},R_{j}\right)=& \int_{O(n)} d\mu(U)d\mu(V)\det\left[1-V^{\dagger}\hat{\gamma} Ur\right]^{-(N+\sigma)}\notag\\
=& \int_{O(n)} d\mu(U)d\mu(V)
\sum_{\lambda}
S_{\lambda}\left(\mathbb{I}_{N_{\sigma}}\right)
S_{\lambda}\left(V^{\dagger}\hat{\gamma} Ur\right)\notag \\
=&\sum_{\lambda}S_{\lambda}\left(\mathbb{I}_{N_{\sigma}}\right)
\int_{O(n)} d\mu(U)d\mu(V)
S_{\lambda}\left(V^{\dagger}\gamma Ur\right)\notag
\end{align}
with $\sigma=-1$ for this ensemble and $\mathbb{I}_{M}$ denotes the identity matrix of dimension $M$. 
The integral over 
$U$ (or $V$) is zero unless the partition is even \cite{McDonalds}.
This means the integers $\lambda=(\lambda_{1},\lambda_{2},\cdots)$ defining the
partition have to be even numbers. This is denoted by $\lambda=2\lambda'=(2\lambda_{1}',2\lambda_{2}',\cdots)$.
Thus the sum over partitions can be written as a sum over even partitions. 
 For even partitions we have through Eq. (\ref{eq:sphericalO})
\begin{align}
 \int_{O(n)}d\mu(U)S_{2\lambda}(AU)=\Omega^{(2)}_{\lambda}(A) \notag
\end{align}
where $\Omega^{(2)}_{\lambda}(A)$ are called the spherical functions defined through their integral property Eq.(\ref{eq:zonalO}).
Thus using Eq. (\ref{eq:zonalO}) yields
\begin{align}
\int_{O(n)} d\mu(U)d\mu(V) S_{2\lambda}\left(V^{\dagger}\hat{\gamma} Ur\right)=&\int_{O(n)} d\mu(V)
\Omega^{(2)}_{\lambda}\left(rV^{\dagger}\hat{\gamma}\right)\notag\\
=&\Omega^{(2)}_{\lambda}\left(\hat{\gamma}\right)\Omega^{(2)}_{\lambda}\left(r \right)\notag
\end{align}
The spherical functions can be expressed in terms of 
Jack Polynomials through Eq. (\ref{eq:spherical}).
We find then 
\begin{align}
\int_{O(n)} dUdV S_{2\lambda}\left(V^{\dagger}\hat{\gamma} Ur\right)
=&
\frac{P^{(2)}_{\lambda}\left(\hat{\gamma}^{2}\right)}{P^{(2)}_{\lambda}(\mathbb{I}_{n})}
\frac{P^{(2)}_{\lambda}\left(r^{2}\right)}{P^{(2)}_{\lambda}(\mathbb{I}_{n})}
\notag
\end{align}
Our integral is then
 \begin{align}
 I_{PRE}\left(\hat{\gamma},R_{j}\right)
=& \sum_{\lambda}S_{2\lambda}(\mathbb{I}_{N_{\sigma}})
\frac{P^{(2)}_{\lambda}\left(\hat{\gamma}^{2}\right)}{P^{(2)}_{\lambda}(\mathbb{I}_{n})}
\frac{P^{(2)}_{\lambda}\left(R\right)}{P^{(2)}_{\lambda}(\mathbb{I}_{n})}
\notag\\
=& \sum_{\lambda}
\frac{e'_{\lambda}(2,N_{\sigma}) }{d'_{\lambda}(2)} P_{\lambda}^{(2)}(\mathbb{I}_{N_{\sigma}})
\frac{P^{(2)}_{\lambda}\left(\hat{\gamma}^{2}\right)}{P^{(2)}_{\lambda}(\mathbb{I}_{n})}
\frac{P^{(2)}_{\lambda}\left(R\right)}{P^{(2)}_{\lambda}(\mathbb{I}_{n})}
\notag 
\end{align}
where we have used Eq. (\ref{eq:schurjack}) to obtain an expression for the 
Schur polynomial evaluated at even partitions.
We can now express our result in terms of Pochhammer symbols using 
Eqs. (\ref{eq:jackidentity}) and (\ref{eq:e}).
\begin{align}
 I_{PRE}\left(\hat{\gamma},R_{j}\right)
 =& \sum_{\lambda}
2^{|\lambda|}\frac{
\left[1+\frac{\left(N_{\sigma}-1\right)}{2}\right]_{\lambda}^{(2)} 
}{d'_{\lambda}(2)} 
\frac{\left[\frac{N_{\sigma}}{2}\right]^{(2)}_{\lambda}}{
\left[\frac{n}{2}\right]^{(2)}_{\lambda}}
\frac{P^{(2)}_{\lambda}\left(\hat{\gamma}^{2}\right)P^{(2)}_{\lambda}\left(R\right)}{P^{(2)}_{\lambda}
(\mathbb{I}_{n})}\label{eq:interimO}
\end{align}
We recognize that Eq. (\ref{eq:interimO}) is  the definition of a Hypergeometric Function of two Matrix Arguments,
$\text{HFMA}_{2}$, Eq. (\ref{eq:HFMArg2}).
There are three types $\text{HFMA}_{2}$, denoted by ${}_{2}\mathcal{F}^{(\alpha)}_{1}\left(a,b;c|X,Y\right)$
 (with index $\alpha=2,1$ or $1/2$) and defined in terms of the symmetric functions as follows: 
\begin{align}
{}_{2}\mathcal{F}^{(\alpha)}_{1}\left(a,b;c|X,Y\right)
=& \sum_{\lambda}
\frac{
\alpha^{|\lambda|}}{d'_{\lambda}(\alpha)} 
\frac{\left[a\right]_{\lambda}^{(\alpha)} 
\left[b\right]^{(\alpha)}_{\lambda}}{
\left[c\right]^{(\alpha)}_{\lambda}}
\frac{P^{(\alpha)}_{\lambda}\left(X\right)P^{(\alpha)}_{\lambda}\left(Y\right)}{P^{(\alpha)}_{\lambda}
(\mathbb{I}_{n})} \notag
\end{align}
Thus we have setting $\sigma=-1$
\begin{align}
 I_{PRE}\left(\hat{\gamma},R_{j}\right)
=& {}_{2} \mathcal{F}_{1}^{(2)}\left(\frac{N}{2},
\frac{N-1}{2};\frac{n}{2} \Big|\hat{\gamma},R\right)
\end{align}
Very little is actually known about determinental/pfaffian representations of
$\text{HFMA}_{2}$. 
If the coupling of the left lead to the quantum dot is independent of the mode
then we are in the case where $\hat{\gamma}=\gamma\mathbb{I}$.
The result reduces then to a $\text{HFMA}_{1}$ , Eq. (\ref{eq:HFMArg1})
\begin{align}
 I_{PRE}\left(\gamma \mathbb{I}_{n},R_{j}\right)
=& {}_{2}F^{(2)}_{1}\left(
\frac{N}{2},
\frac{N-1}{2};\frac{n}{2}\Big|\gamma^{2}R\right)
\label{eq:HFMAO}
\end{align}
Before turning to the question of
representations of the $\text{HFMA}_{1}$ we analyze the PQE case in the same manner. 
The results will be HFMA$_{1,2}$ with the index $\alpha=1/2$.
\subsection{Poisson Quaternion Ensemble}
For quaternion ensemble we need to perform the following integration
\begin{align}
 I_{PQE}\left(\hat{\gamma},R_{j}\right)
&= \int_{Sp(n)} d\mu(U)d\mu(V)\left|\det\left[1-\hat{\gamma}UrV\right]\right|^{-N_{\sigma}}
\end{align}
Using the fact that the unitary matrices are symplectic we have  $
U^{\dagger}= U^{R}=-ZU^{T}Z$
with $ Z=\mathbb{I}_{n} \otimes i\tau_{y}$ and $\tau_{y}$ the Pauli matrix, 
we gather that the determinant is real even though the matrix is complex. 
\begin{align}
 \det\left[1-\hat{\gamma} UrV\right]^{*}
=& \det\left[1-\hat{\gamma} U^{*}rV^{*}\right]\notag \\
=& \det\left[1-\hat{\gamma} ZUZrZVZ\right]\notag \\
=& \det\left[1-\hat{\gamma} UrV\right]\notag 
\end{align}
In the last step we have used the fact that $r$ has a double degeneracy,
$r=\text{diag}\left\{r_{1},\cdots r_{n}\right\}\otimes \mathbb{I}_{2\times 2}$.
Thus for the PQE we have
\begin{align}
 I_{PQE}\left(\hat{\gamma},R_{j}\right)
&= \int_{Sp(n)}d\mu(U)d\mu(V)\det\left[1-\hat{\gamma} UrV\right]^{-N_{\sigma}} \notag \\
&=
\sum_{\lambda}
\frac{\left[N_{\sigma}\right]_{\lambda}^{(1)}}{h_{\lambda}(1)}
\int_{Sp(n)} d\mu(U)d\mu(V)
S_{\lambda}\left(\hat{\gamma} UrV\right)\notag 
\end{align}
Similarly to the PRE the integral will be zero for partitions which do not have 
a specific form , namely the form $\lambda=\lambda' \cup \lambda'$ for any $\lambda'$.
The partition $\lambda' \cup \lambda'$ is defined by having each 
integer twice.
That is $\lambda' \cup \lambda'=(\lambda_{1}',\lambda_{1}',\lambda_{2}',\lambda_{2}',\cdots)$.
We have then 
\begin{align}
 I_{PQE}\left(\hat{\gamma},R_{j}\right)
&=
\sum_{\lambda}
\frac{\left[N_{\sigma}\right]_{\lambda \cup \lambda}^{(1)}}{h_{\lambda\cup \lambda}(1)}
\int_{Sp(n)} d\mu(U)d\mu(V)
S_{\lambda\cup \lambda}\left(\hat{\gamma} UrV\right)\notag 
\end{align}
For partitions which do  have this form we can perform the integrations
using Eqs. (\ref{eq:sphericalS}), (\ref{eq:zonalS}) and (\ref{eq:spherical})
\begin{align}
 \int_{Sp(n)} d\mu(U)d\mu(V)
S_{\lambda \cup \lambda}\left(\hat{\gamma} UrV\right)
&=
\int_{Sp(n)} d\mu(V)
\Omega^{(1/2)}_{\lambda }\left(rV\hat{\gamma}\right)\notag \\
&=\Omega^{(1/2)}_{\lambda }\left(\hat{\gamma}\right)
\Omega^{(1/2)}_{\lambda }\left(r\right)
\notag \\
&=
\frac{P^{(1/2)}_{\lambda}\left(\hat{\gamma}^{2}\right)}{P^{(1/2)}_{\lambda}(\mathbb{I}_{n})}
\frac{P^{(1/2)}_{\lambda}\left(R\right)}{P^{(1/2)}_{\lambda}(\mathbb{I}_{n})}
\notag
\end{align}
This leads to the following expression : 
\begin{align}
 I_{PQE}\left(\hat{\gamma},R_{j}\right)
&=
\sum_{\lambda}
 \frac{e'_{\lambda}\left(\frac{1}{2},\frac{N_{\sigma}}{2}\right) b_{\lambda}\left(\frac{1}{2},\frac{N_{\sigma}}{2}\right) }{
 d'_{\lambda}\left(\frac{1}{2}\right)h_{\lambda}\left(\frac{1}{2}\right)}
\frac{P^{\left(\frac{1}{2}\right)}_{\lambda}\left(\hat{\gamma}^{2}\right)}{P^{\left(\frac{1}{2}\right)}_{\lambda}(\mathbb{I}_{n})}
\frac{P^{\left(\frac{1}{2}\right)}_{\lambda}\left(R\right)}{P^{\left(\frac{1}{2}\right)}_{\lambda}(\mathbb{I}_{n})}
\notag
\end{align}
where we have used Eq. (\ref{eq:identityquaternion}). 
Using Eqs. (\ref{eq:e} ) and (\ref{eq:jackidentity}) we have 
\begin{align}
 I_{PQE}\left(\hat{\gamma},R_{j}\right)
&=
\sum_{\lambda}\left(\frac{1}{2}\right)^{|\lambda|}
 \frac{ 
 \left[N_{\sigma}-1\right]_{\lambda}^{(1/2)}
 \left[N_{\sigma}\right]_{\lambda}^{(1/2)}
  }{
 d'_{\lambda}\left(\frac{1}{2}\right)
  \left[2n\right]_{\lambda}^{(1/2)}
  }
\frac{
P^{\left(\frac{1}{2}\right)}_{\lambda}\left(\hat{\gamma}^{2}\right)
P^{\left(\frac{1}{2}\right)}_{\lambda}\left(R\right)
}{
P^{\left(\frac{1}{2}\right)}_{\lambda}(\mathbb{I}_{n})
}
\notag
\end{align}
we identify this solution with the $\text{HFMA}_{2}$ the index with $\alpha=1/2$. 
Setting $\sigma=1$ we gather
\begin{align}
 I_{PQE}\left(\hat{\gamma},R_{j}\right)
&= {}_{2}\mathcal{F}^{\left(\frac{1}{2}\right)}_{1}\left(N,
N+1;2n\Big| \hat{\gamma}^{2}, R\right)\notag\\
&= {}_{2}\mathcal{F}^{\left(\frac{1}{2}\right)}_{1}\left(2(n+m),
2(n+m)+1;2n\Big| \hat{\gamma}^{2}, R\right)
\end{align}
where we have set $\sigma=1$
and the case of $\hat{\gamma}$ proportional to the identity yields 
\begin{align}
 I_{PQE}\left(\gamma \mathbb{I},R_{j}\right)
&= {}_{2}F^{\left(\frac{1}{2}\right)}_{1}\left(N,
N+1 ;2n\Big|\gamma^{2} R\right) \notag\\
&= {}_{2}F^{\left(\frac{1}{2}\right)}_{1}\left(2(n+m),
2(n+m)+1 ;2n\Big|\gamma^{2} R\right)
\label{eq:HFMAS}
\end{align}
For the two ensembles we have then the following j.p.d.f.
when $\hat{\gamma}$ is arbitrary
\begin{align}
 \mathcal{P}_{\alpha}(R_{j})
 =\frac{1}{\mathcal{C}(\hat{\gamma})}\prod_{j=1}^{n}\left(R_{j}-1\right)^{\frac{1}{\alpha}\left(m-n+1\right)-1}
 R_{j}^{\frac{\eta}{2}}|\Delta\left(R_{j}\right)|^{\frac{2}{\alpha}}
{}_{2}\mathcal{F}_{1}^{(\alpha)}\left(\frac{m+n}{\alpha},
\frac{m+n}{\alpha}+\frac{\eta}{2};\frac{n}{\alpha}\Big|\hat{\gamma}^{2},R\right),
\label{eq:j.p.d.f.1}
\end{align}
and when $\hat{\gamma}=\gamma\mathbb{I}$ it simplifies to 
\begin{align}
 \mathcal{P}_{\alpha}(R_{j})
 =
 \frac{1}{\mathcal{C}(\hat{\gamma})}
 \prod_{j=1}^{n}\left(R_{j}-1\right)^{\frac{1}{\alpha}\left(m-n+1\right)-1}
 R_{j}^{\frac{\eta}{2}}|\Delta\left(R_{j}\right)|^{\frac{2}{\alpha}}
{}_{2}F_{1}^{(\alpha)}\left(\frac{m+n}{\alpha},
\frac{m+n}{\alpha}+\frac{\eta}{2};\frac{n}{\alpha}\Big|\gamma^{2}R\right) \label{eq:j.p.d.f.}
\end{align}
where we have added the index $\alpha$ to the j.p.d.f. of Eq. (\ref{eq:jpdf})
that specifies the ensemble. $\alpha=2$ for the real ensemble
and $\frac{1}{2}$ for the quaternion one.
$\mathcal{C}$ denotes the normalization constant which we compute now.
\section{Normalization}
From Eq. (\ref{eq:j.p.d.f.1}) we gather the normalization constant is given by the following integral
\begin{align}
\mathcal{C}(\hat{\gamma})
=&\prod_{j=1}^{n}
 \int_{0}^{1}dR_{j}
 \left(R_{j}-1\right)^{\frac{1}{\alpha}\left(m-n+1\right)-1}
 R_{j}^{\frac{\eta}{2}}|\Delta\left(R_{j}\right)|^{\frac{2}{\alpha}}
{}_{2}\mathcal{F}_{1}^{(\alpha)}\left(\frac{m+n}{\alpha},
\frac{m+n}{\alpha}+\frac{\eta}{2};\frac{n}{\alpha}\Big|\hat{\gamma}^{2};R\right) \notag\\
=& \sum_{\lambda}
\frac{
\alpha^{|\lambda|}}{d'_{\lambda}(\alpha)} 
\frac{\left[\frac{m+n}{\alpha}\right]_{\lambda}^{(\alpha)} 
\left[\frac{m+n}{\alpha}+\frac{\eta}{2}\right]^{(\alpha)}_{\lambda}}{
\left[\frac{n}{\alpha}\right]^{(\alpha)}_{\lambda}}
\frac{
P^{(\alpha)}_{\lambda}\left(\hat{\gamma}^{2}\right)
}{P^{(\alpha)}_{\lambda}\left(\mathbb{I}_{n}\right)
}
\prod_{j=1}^{n}
 \int_{0}^{1}dR_{j}
 \left(R_{j}-1\right)^{\frac{1}{\alpha}\left(m-n+1\right)-1}
 R_{j}^{\frac{\eta}{2}}|\Delta\left(R_{j}\right)|^{\frac{2}{\alpha}}
P^{(\alpha)}_{\lambda}\left(R\right) \notag
\end{align}
Given the Selberg integral over Jack polynomials
\begin{align}
\prod_{j=1}^{n} \int_{0}^{1}dR_{j}
 \left(1-R_{j}\right)^{y}
 R_{j}^{x}|\Delta\left(R_{j}\right)|^{\frac{2}{\alpha}}P^{(\alpha)}_{\lambda}\left(R\right)
=&P^{(\alpha)}_{\lambda}\left(\mathbb{I}_{n}\right)
\frac{\left[x+1+\frac{n-1}{\alpha}\right]^{\alpha}_{\lambda}}{
\left[x+y+2+\frac{2}{\alpha}\left(n-1\right)\right]^{\alpha}_{\lambda}}
S_{n}(x,y,\alpha)
\notag\\
S_{n}(x,y,\alpha)
=&
\prod_{j=1}^{n} \int_{0}^{1}dR_{j}
 \left(1-R_{j}\right)^{y}
 R_{j}^{x}|\Delta\left(R_{j}\right)|^{\frac{2}{\alpha}}\notag\\
=&\prod_{j=0}^{n-1}\frac{
\Gamma\left(x+1+\frac{j}{\alpha}\right)
\Gamma\left(y+1+\frac{j}{\alpha}\right)
\Gamma\left(1+\frac{j}{\alpha}\right)}{
\Gamma\left(x+y+2+\frac{n+j-1}{\alpha}\right)
\Gamma\left(1+\frac{1}{\alpha}\right)
} 
\end{align}
we have
\begin{align}
\prod_{j=1}^{n} \int_{0}^{1}dR_{j}
 \left(R_{j}-1\right)^{\frac{1}{\alpha}\left(m-n+1\right)-1}
 R_{j}^{\frac{\eta}{2}}|\Delta\left(R_{j}\right)|^{\frac{2}{\alpha}}P^{(\alpha)}_{\lambda}\left(R\right) 
=&C_{n}P^{(\alpha)}_{\lambda}\left(\mathbb{I}_{n}\right)
\frac{\left[\frac{\eta}{2}+1-\frac{1}{\alpha}+\frac{n}{\alpha}\right]^{\alpha}_{\lambda}}{
\left[\frac{\eta}{2}+1-\frac{1}{\alpha}+\frac{1}{\alpha}\left(m+n\right)\right]^{\alpha}_{\lambda}} \notag \\
C_{n}=&
S_{n}\left(
\frac{\eta}{2},\frac{1}{\alpha}\left(m-n+1\right)-1
,\alpha\right)
\end{align}
 $C_{n}$ is the normalization constant for the circular ensemble ($\hat{\gamma}=0$). 
Given that $\frac{\eta}{2}+1-\frac{1}{\alpha}=0$ for both ensembles we have 
\begin{align}
 \int_{0}^{1}dR_{j}
 \left(R_{j}-1\right)^{\frac{1}{\alpha}\left(m-n+1\right)-1}
 R_{j}^{\frac{\eta}{2}}|\Delta\left(R_{j}\right)|^{\frac{2}{\alpha}}P^{(\alpha)}_{\lambda}\left(R\right)
=&C_{n}
P^{(\alpha)}_{\lambda}\left(\mathbb{I}_{n}\right)
\frac{\left[\frac{n}{\alpha}\right]^{\alpha}_{\lambda}}{
\left[\frac{1}{\alpha}\left(m+n\right)\right]^{\alpha}_{\lambda}}
\end{align}
and so 
\begin{align}
\mathcal{C}(\hat{\gamma})
=&C_{n}
 \sum_{\lambda}
\frac{
\alpha^{|\lambda|}}{d'_{\lambda}(\alpha)} 
\left[\frac{m+n}{\alpha}+\frac{\eta}{2}\right]^{(\alpha)}_{\lambda}
P^{(\alpha)}_{\lambda}\left(\hat{\gamma}^{2}\right)\notag\\
=&C_{n}
\prod_{j=1}^{n}\left[1-\gamma^{2}_{j}\right]^{-\left(\frac{m+n}{\alpha}+\frac{\eta}{2}\right)} \notag \\
=&C_{n}
\det\left[1-\hat{\gamma}^{2}\right]^{-\left(\frac{m+n}{\alpha}+\frac{\eta}{2}\right)}
\end{align}
For the orthogonal ensembles this yields
$\det\left[1-\hat{\gamma}^{2}\right]^{-\left(\frac{N}{2}-\frac{1}{2}\right)}$
and for the quaternion we have 
$\det\left[1-\hat{\gamma}^{2}\otimes \mathbb{I}_{2}\right]^{-\left(\frac{N}{2}+\frac{1}{2}\right)}$

\section{Representations of $\text{HFMA}_{1}$}
\label{sec:representationHFMA}
In this section we will derive another integral representation for the $\text{HFMA}_{1}$.
We will show that the following matrix integral
\begin{align}
F_{a,b}^{p,\alpha}\left(X\right)
 =&
 \frac{1}{Z_{p}(a,b)}
 \prod_{j=1}^{p} \int_{0}^{\infty}
 dy_{j}\left|\Delta\left(y_{j}\right)\right|^{\frac{2}{\alpha}} 
\frac{y_{j}^{a} }{
\left(1+y_{j}\right)^{b }} \prod_{k,j}^{n,p}\left(1+x_{k}y_{j} \right),
\end{align}
with $\alpha=1/2,1,2$ and $Z_{p}(a,b)$ the normalization constant
\begin{align}
Z_{p}(a,b)=&
 \prod_{j=1}^{p} \int_{0}^{\infty}
 dy_{j}\left|\Delta\left(y_{j}\right)\right|^{\frac{2}{\alpha}} 
\frac{y_{j}^{a} }{
\left(1+y_{j}\right)^{b }}
\notag \\
=& 2^{2(p-1)\left(1+\frac{1}{\alpha}(p-1)\right)}
\prod_{j=0}^{p-1}\frac{
\Gamma\left(a+1+\frac{j}{\alpha}\right)
\Gamma\left(b-a-1-\frac{2}{\alpha}(p-1)+\frac{j}{\alpha}\right)
\Gamma\left(1+\frac{j}{\alpha}\right)}{
\Gamma\left(b-\frac{1}{\alpha}(p-1)+\frac{j}{\alpha}\right)
\Gamma\left(1+\frac{1}{\alpha}\right)
} 
\end{align}
is
a $\text{HFMA}_{1}$.
We first use the dual Cauchy identity
\begin{align}
\prod_{j,k}^{n,p} \left(1+x_{j} y_{k} \right)
=&
\sum_{\lambda}
P_{\lambda^{t}}^{\left(\frac{1}{\alpha}\right)}\left(X\right)
P_{\lambda}^{(\alpha)}\left(Y\right) \notag
\end{align}
leading to 
\begin{align}
F_{a,b}^{p,\alpha}\left(X\right)
 =&\frac{1}{Z_{p}(a,b)}
\sum_{\lambda} 
P_{\lambda^{t}}^{\left(\frac{1}{\alpha}\right)}\left(X\right)
 \prod_{j=1}^{p} \int_{0}^{\infty}
 dy_{j}\left|\Delta\left(y_{j}\right)\right|^{\frac{2}{\alpha}} 
\frac{y_{j}^{a}}{ \left(1+y_{j}\right)^{b}}
P_{\lambda}^{(\alpha)}\left(y_{j}\right)
\label{eq:F1}
\end{align}
The sum over $\lambda$ is over partitions such that $l(\lambda)\leq p$
and $l(\lambda^{t})\leq n$.
The dual generalized Selberg integrals states the following identity holds
\begin{align}
 \prod_{j=1}^{p} \int_{0}^{\infty}
 dy_{j}\left|\Delta\left(y_{j}\right)\right|^{\frac{2}{\alpha}} 
\frac{y_{j}^{a}}{ \left(1+y_{j}\right)^{b}}
P_{\lambda}^{(\alpha)}\left(y_{j}\right)
=
Z_{p}(a,b)
P_{\lambda}^{(\alpha)}\left(\mathbb{I}_{p}\right)
\frac{\left[a+1+\frac{p-1}{\alpha}\right]^{(\alpha)}_{\lambda}}{(-1)^{|\lambda|}\left[a+2+2\frac{p-1}{\alpha}-b\right]^{(\alpha)}_{\lambda}}
\notag
\end{align}
provided $l(\lambda^{t})< b-a-1-2\frac{p-1}{\alpha}$. 
Since the sum is over partitions such that $l(\lambda^{t})\leq n$,
the condition is fulfilled for all partitions if  $n<b-a-1-2\frac{p-1}{\alpha}$. Let us assume this last 
inequality holds. Using the dual generalized Selberg integral in Eq. (\ref{eq:F1}) we have  
\begin{align}
F_{a,b}^{p,\alpha}\left(X\right)
 =&
\sum_{\lambda}
\frac{\left[a+1+\frac{p-1}{\alpha}\right]^{(\alpha)}_{\lambda}}{(-1)^{|\lambda|}\left[a+2+2\frac{p-1}{\alpha}-b\right]^{(\alpha)}_{\lambda}}
P_{\lambda^{t}}^{\left(\frac{1}{\alpha}\right)}\left(X\right)
P_{\lambda}^{(\alpha)}\left(\mathbb{I}_{p}\right)
\label{eq:sum}
\end{align}
We would like to rewrite this expression solely in terms
of $\lambda^{t}$ and $\frac{1}{\alpha}$ so as 
to compare it with the definition of $\text{HFMA}_{1}$.
For the Jack polynomial evaluated at identity 
we have 
\begin{align}
P_{\lambda}^{(\alpha)}\left(\mathbb{I}_{p}\right)
=\frac{\alpha^{|\lambda|}\left[\frac{p}{\alpha}\right]_{\lambda}^{(\alpha)}}{
h_{\lambda}(\alpha)} 
\end{align}
and using the following relationship between Pochhammer symbols of different index $\alpha$
\begin{align}
\left[s\right]^{(\alpha)}_{\lambda}
 =&
 \frac{ (-1)^{|\lambda|}}{\alpha^{|\lambda|}}\left[-\alpha s\right]^{(\frac{1}{\alpha})}_{\lambda^{t}}\label{eq:Pochhammerdual}
\end{align}
we can express the Jack polynomial evaluated at the identity $P_{\lambda}^{(\alpha)}\left(\mathbb{I}_{p}\right)$
as 
\begin{align}
P_{\lambda}^{(\alpha)}\left(\mathbb{I}_{p}\right)
=\frac{
  (-1)^{|\lambda|}\left[-p\right]^{(\frac{1}{\alpha})}_{\lambda^{t}}
}{
h_{\lambda}(\alpha)} \label{eq:jackidentitydual}
\end{align}
Using Eq. (\ref{eq:Pochhammerdual}) we can also rewrite the 
ratios of Pochhammer symbols in the sum Eq.(\ref{eq:sum}) as
\begin{align}
\frac{\left[a+1+\frac{p-1}{\alpha}\right]^{(\alpha)}_{\lambda}}{
\left[a+2+2\frac{p-1}{\alpha}-b\right]^{(\alpha)}_{\lambda}}
=&
\frac{\left[-\alpha\left(a+1+\frac{p-1}{\alpha}\right)\right]^{\left(\frac{1}{\alpha}\right)}_{\lambda^{t}}}{
\left[-\alpha\left(a+2+2\frac{p-1}{\alpha}-b\right)\right]^{\left(\frac{1}{\alpha}\right)}_{\lambda^{t}}}\label{eq:ratiosdual}
\end{align}
In addition we have the following relations
\begin{align}
h_{\lambda}(\alpha)
=\alpha^{|\lambda|}d'_{\lambda^{t}}\left(\frac{1}{\alpha}\right)\label{eq:hooklengthdual}
\end{align}
Combining Eqs. (\ref{eq:jackidentitydual}), (\ref{eq:ratiosdual}) and (\ref{eq:hooklengthdual}) in Eq. (\ref{eq:sum})
we gather
\begin{align}
F_{a,b}^{p,\alpha}\left(X\right)
 =&
\sum_{\lambda;l(\lambda)\leq p;
l(\lambda^{t})\leq n}
\frac{\left[-p\right]^{(\frac{1}{\alpha})}_{\lambda^{t}}}{
\alpha^{|\lambda|}d'_{\lambda^{t}}\left(\frac{1}{\alpha}\right)}
\frac{\left[-\alpha\left(a+1\right)+1-p\right]^{\left(\frac{1}{\alpha}\right)}_{\lambda^{t}}}{
\left[-\alpha\left(a+2\right)+2(1-p)+\alpha b\right]^{\left(\frac{1}{\alpha}\right)}_{\lambda^{t}}}
P_{\lambda^{t}}^{\left(\frac{1}{\alpha}\right)}\left(X\right)
\end{align}
Since there is a one to one correspondence between partitions and their conjugates, summing over all partitions
is the same as summing over all conjugate partitions.
We make the change in notation $\lambda^{t}\rightarrow \lambda$
and denote $\alpha'=\frac{1}{\alpha}$ leading to
\begin{align}
F_{a,b}^{p,\alpha}\left(X\right)
=&
\sum_{\lambda;l(\lambda^{t})\leq p;
l(\lambda)\leq n} 
\frac{(\alpha')^{|\lambda|}
\left[-p\right]^{(\alpha')}_{\lambda}
\left[-\frac{1}{\alpha'}\left(a+1\right)+1-p\right]^{\left(\alpha'\right)}_{\lambda}}{
d'_{\lambda}\left(\alpha'\right)
\left[-\frac{1}{\alpha'}\left(a+2\right)+2(1-p)+\frac{ b}{\alpha'}\right]^{\left(\alpha'\right)}_{\lambda}}
P_{\lambda}^{\left(\alpha'\right)}\left(X\right) 
\end{align}
The Pochhammer symbol
$\left[-p\right]^{(\alpha')}_{\lambda}$ is zero if $\lambda_{1}>p$. Since
$\lambda_{1}=l(\lambda^{t})$
 the restriction $l(\lambda^{t})\leq p$ is automatically satisfied in the sum.
Thus we have 
\begin{align}
&
F_{a,b}^{p,\alpha}\left(X\right)
=
\sum_{\lambda;l(\lambda)\leq n}
\frac{
(\alpha')^{|\lambda|}
\left[-p\right]^{(\alpha')}_{\lambda}
\left[-q
\right]^{\left(\alpha'\right)}_{\lambda}
}{
d'_{\lambda}\left(\alpha'\right)
\left[c
\right]^{\left(\alpha'\right)}_{\lambda}}P_{\lambda}^{\left(\alpha'\right)}\left(X\right) 
 \label{eq:HFMAsum}
\end{align}
with
\begin{align}
-q=&-\frac{1}{\alpha'}\left(a+1\right)+1-p \\
c=&-\frac{1}{\alpha'}\left(a+2\right)+2(1-p)+\frac{ b}{\alpha'}
\end{align}
The sum in Eq. (\ref{eq:HFMAsum}) is known to be a $\text{HFMA}_{1}$ , Eq. (\ref{eq:HFMArg1}).
\begin{align}
 F_{a,b}^{p,\alpha}\left(X\right)
= _{2}F_{1}^{(\alpha')}\left(-p,-q;c
\big|X
\right)
\end{align}
This identity holds subjected to the condition
which came from the use of the dual Selberg integral.
\begin{align}
 n< b-a-1-2\alpha'(p-1) \notag
\end{align}
Thus for the $\text{HFMA}_{1}$
$ _{2}F_{1}^{(\alpha')}\left(-p,-q;c
\big|X
\right)$
the condition translates into 
(with $c=\frac{1}{\alpha'}\left(b-a-2\right)-2(p-1)$)
\begin{align}
\frac{ n-1}{\alpha'}&< c
\end{align}
If the condition is met the $\text{HFMA}_{1}$ has the  following integral
representation
\begin{align}
{}_{2}F_{1}^{(\alpha')}\left(-p,-q;c|X\right) 
 =&\frac{1}{Z_{p}}
 \prod_{j=1}^{p} \int_{0}^{\infty}
 dy_{j}\left|\Delta\left(y_{j}\right)\right|^{\frac{2}{\alpha}} 
\frac{ y_{j}^{\alpha'\left(q-p+1\right)-1} }{
\left(1+y_{j}\right)^{\alpha'\left(c+p+q-1\right)+1 }}
 \prod_{k,j}^{n,p}\left(1+x_{k}y_{j} \right)
\notag \\
Z_{p}=&
 \prod_{j=1}^{p} \int_{0}^{\infty}
 dy_{j}\left|\Delta\left(y_{j}\right)\right|^{\frac{2}{\alpha}} 
\frac{ y_{j}^{\alpha'\left(q-p+1\right)-1} }{
\left(1+y_{j}\right)^{\alpha'\left(c+p+q-1\right)+1 }}\notag\\
=& 2^{2(p-1)\left(1+\alpha'(p-1)\right)}
\prod_{j=0}^{p-1}\frac{
\Gamma\left(\alpha'\left(q-j\right)\right)
\Gamma\left(\alpha'(c+j)
+1
\right)
\Gamma\left(1+j\alpha'\right)}{
\Gamma\left(\alpha'\left(c+q+j\right)+1\right)
\Gamma\left(1+\alpha'\right)
} 
\end{align}
Performing the change of variables $y_{j}=\frac{1+\lambda_{j}}{1-\lambda_{j}}$ we have 
\begin{align}
&{}_{2}F_{1}^{(\alpha')}\left(-p,-q;c|X\right) \notag\\
=&
\frac{
1}{\tilde{Z}_{p}}
 \prod_{k}^{n}(1-x_{k})^{p}
\prod_{j=1}^{p} \int_{-1}^{1}
 d\lambda_{j}\left|\Delta\left(\lambda_{j}\right)\right|^{\frac{2}{\alpha}} 
 \left(1+\lambda_{j}\right)^{\alpha'\left(q-p+1\right)-1} 
\left(1-\lambda_{j}\right)^{\alpha'c-n}
\prod_{k,j}^{n,p}\left(\frac{1+x_{k}}{1-x_{k}}-\lambda_{j} \right)\label{eq:HFMA1Jacobi} \\
\tilde{Z}_{p}=&\frac{Z_{p}}{2^{(p-1)\left(1+p\alpha'\right)-\alpha'(c+p+q-1)}} \notag
\end{align}
and if we set $z_{k}=\frac{1+x_{k}}{1-x_{k}}$ then the integral is the average of a product of characteristic polynomials
of a Jacobi Ensemble, when $\frac{2}{\alpha}=1,2$ and the average of a product of square roots of characteristic polynomials
when $\frac{2}{\alpha}=4$ .

\section{Representation of the j.p.d.f.}
\label{sec:pfaffianrep}
The representation of HFMA$_{1}$ derived above applies only for negative integer values  of the first two
arguments $-p,-q$ while in Eq. (\ref{eq:j.p.d.f.}) the first two
arguments of the HFMA$_{1}$ has clearly positive values.
To obtain a representation of the j.p.d.f. we use the well known Kummer's relations for the $\text{HFMA}_{1}$,
\begin{align}
{}_{2} F_{1}^{(\alpha)}\left(a,b;c\Big|X\right)
= \frac{{}_{2} F_{1}^{(\alpha)}\left(c-a,c-b;c\Big|X\right)}{
\det\left[1-X\right]^{a+b-c}}
\end{align}
We first look at the PQE , $\alpha=\frac{1}{2}$.
For the PQE we have from Eq. (\ref{eq:HFMAS})
\begin{align}
I_{PQE}\left(\gamma \mathbb{I},R\right) 
&= 
\frac{{}_{2}F^{(\frac{1}{2})}_{1}\left(
-2m,-2m-1
;2n\Big|\gamma^{2} R\right)}{
\det\left[1-\gamma^{2} R\right]^{
4m+2n+1
}
}
\notag
\end{align}
In Eq. (\ref{eq:HFMA1Jacobi}) we set 
$p=\frac{m}{\alpha}$and $q=\frac{m}{\alpha}+\frac{\eta}{2}$,
and  have then 
$p=2m$, $q=2m+1$.
For these values of $p$, $q$ and $c$ the weight in 
 (\ref{eq:HFMA1Jacobi}) simplifies to $ \left(1+\lambda_{j}\right)^{\alpha\left(q-p+1\right)-1} 
\left(1-\lambda_{j}\right)^{\alpha c-n}
=1$.
The HFMA$_{1}$ appearing can be expressed as a pfaffian over a Vandermonde determinant
through Eqs.(\ref{eq:pfaffian}),(\ref{eq:pfaffian2}) depending on whether $n$ is even or odd.
We assume $n$ is even and using the result of Eq. (\ref{eq:pfaffian2}) we gather 
\begin{align}
{}_{2}F^{(\frac{1}{2})}_{1}\left(
-2m,-2m-1
;2n\Big|\gamma^{2} R\right)
&=\frac{1}{
\tilde{Z}_{2m}}
 \prod_{k=1}^{n}(1-\gamma^{2}R_{k})^{2m}
\prod_{j=1}^{2m} \int_{-1}^{1}
 d\lambda_{j}\left|\Delta\left(\lambda_{j}\right)\right|
 \prod_{k,j}^{n,2m}\left(\frac{1+\gamma^{2}R_{k}}{1-\gamma^{2}R_{k}}-\lambda_{j} \right)\notag\\
&=\frac{1}{
\tilde{Z}_{2m}}
 \prod_{k=1}^{n}(1-\gamma^{2}R_{k})^{2m}
\frac{1}{\Delta\left(
\frac{1+\gamma^{2}R_{k}}{1-\gamma^{2}R_{k}}\right)}
\text{Pf}_{j,k\leq n}\left[f_{jk}
\right]\notag\\
&=\frac{1}{
\tilde{Z}_{2m}
(2\gamma^{2})^{\frac{n(n-1)}{2}}
}
 \prod_{k=1}^{n}(1-\gamma^{2}R_{k})^{2m+n-1}
\frac{1
}{\Delta\left(R_{k}\right)}
\text{Pf}_{j,k\leq n}\left[f_{jk}
\right]\notag
\end{align}
with
\begin{align}
\tilde{Z}_{2m}=& 
2^{2m^{2}+2+n}
 \prod_{j=0}^{2m-1}\frac{
\Gamma\left(\frac{1}{2}\left(2m+1-j\right)\right)
\Gamma\left(\frac{1}{2}(2n+j)
+1
\right)
\Gamma\left(1+j\frac{1}{2}\right)}{
\Gamma\left(\frac{1}{2}\left(2n+2m+1+j\right)+1\right)
\Gamma\left(1+\frac{1}{2}\right)
} \notag
\end{align}
and the entries $f_{jk}$ are given in terms of the Jacobi skew orthogonal polynomials by Eqs. (\ref{eq:fjk2}) with the weight $w(u)=1$.
The arguments $v_{k}$ in Eqs. (\ref{eq:fjk2}) are $\frac{1+\gamma^{2}R_{k}}{1-\gamma^{2}R_{k}}$. 
The j.p.d.f. is then for the PQE  
\begin{align}
 \mathcal{P}_{\alpha=\frac{1}{2}}(R_{j})=&
 \frac{1}{C(\gamma)}
 \Delta\left(R_{j}\right)^{3}
\prod_{j=k}^{n}\frac{ R_{k}^{\frac{\eta}{2}}
 \left(1-R_{k}\right)^{2\left(m-n\right)+1}
}{
\left(1-\gamma^{2} R_{k}\right)^{
2m+n+2
}
}
\text{Pf}\left[f_{jk}
\right]\\
C(\gamma)=&  \mathcal{C}(\hat{\gamma})
\tilde{Z}_{2m}(2\gamma^{2})^{n\frac{(n-1)}{2}} \notag
\end{align}
Given the antisymmetry of the Vandermonde determinant and the pfaffian under exchange 
of two variables $R_{j}$ and $R_{k}$ we gather
\begin{align}
 \mathcal{P}_{\alpha=\frac{1}{2}}(R_{j})=&
\frac{1}{C(\hat{\gamma})} \Delta\left(R_{j}\right)^{3}
\prod_{k=1}^{n}\frac{ R_{k}^{\frac{\eta}{2}}
 \left(1-R_{k}\right)^{2\left(m-n\right)+1}
}{
\left(1-\gamma^{2} R_{k}\right)^{
2m+n+2
}
}
\prod_{s=1}^{\frac{n}{2}}
F\left(\frac{1+\gamma^{2}R_{2s-1}}{1-\gamma^{2}R_{2s-1}},\frac{1+\gamma^{2}R_{2s}}{1-\gamma^{2}R_{2s}}
\right)
\end{align}
where the function $F(u,v)$ is given by Eq. (\ref{eq:Fodd}).
For $\alpha=2$ ($\beta=1$, the PRE case) we have from Eq. (\ref{eq:HFMAO})
\begin{align}
I_{PRE}\left(\gamma\mathbb{I}, R\right) 
&= 
\frac{{}_{2}F^{(2)}_{1}\left(
-\frac{m}{2},-\frac{m-1}{2}
;\frac{n}{2}\Big|\gamma^{2} R\right)}{
\det\left[1-\gamma^{2} R\right]^{
\frac{2m+n-1}{2}
}
}
\notag
\end{align}
There are two possibilities, $m$ even or odd. For $m$ even we take
$p=\frac{m}{2}$ , $q=\frac{m-1}{2}$. 
For these values of $p$, $q$ and $c$ the weight in 
 (\ref{eq:HFMA1Jacobi}) simplifies to $ \left(1+\lambda_{j}\right)^{\alpha\left(q-p+1\right)-1} 
\left(1-\lambda_{j}\right)^{\alpha c-n}
=1$.
\begin{align}
{}_{2}F^{(2)}_{1}\left(
-\frac{m}{2},-\frac{m-1}{2}
;\frac{n}{2}\Big|\gamma^{2} R\right)
&=\frac{1}{
\tilde{Z}_{\frac{m}{2}}}
 \prod_{k=1}^{n}(1-\gamma^{2}R_{k})^{\frac{m}{2}}
\prod_{j=1}^{\frac{m}{2}} \int_{-1}^{1}
 d\lambda_{j}\left|\Delta\left(\lambda_{j}\right)\right|^{4}
 \prod_{k,j}^{n,\frac{m}{2}}\left(\frac{1+\gamma^{2}R_{k}}{1-\gamma^{2}R_{k}}-\lambda_{j} \right)
\end{align}
 For $m$ odd we take
$p=\frac{m-1}{2}$ , $q=\frac{m}{2}$. 
For these values of $p$, $q$ and $c$ the weight in 
 (\ref{eq:HFMA1Jacobi}) simplifies to 
 $ 
 \left(1+\lambda_{j}\right)^{\alpha\left(q-p+1\right)-1} 
\left(1-\lambda_{j}\right)^{\alpha c-n}
= \left(1+\lambda_{j}\right)^{2}$.
\begin{align}
&{}_{2}F^{(2)}_{1}\left(
-\frac{m}{2},-\frac{m-1}{2}
;\frac{n}{2}\Big|\gamma^{2} R\right) \notag\\
&=\frac{1}{
\tilde{Z}_{\frac{m-1}{2}}}
 \prod_{k=1}^{n}(1-\gamma^{2}R_{k})^{\frac{m-1}{2}}
\prod_{j=1}^{\frac{m-1}{2}} \int_{-1}^{1}
 d\lambda_{j}\left|\Delta\left(\lambda_{j}\right)\right|^{4}(1+\lambda_{k})^{2}
 \prod_{k,j}^{n,\frac{m-1}{2}}\left(\frac{1+\gamma^{2}R_{k}}{1-\gamma^{2}R_{k}}-\lambda_{j} \right)
\end{align}
The products appearing in the average here are not characteristic polynomials but rather 
square roots of characteristic polynomials and this is why we can not follow the same type of calculation as for the 
PQE.

\section{Conclusion}
We have shown that when considering
thermal transport through a semi-non-ideal Andreev quantum dot 
the j.p.d.f. is related to 
Hypergeometric Functions of Matrix Argument quite analogously to the 
case of electric transport studied in \cite{VidalKanzieper} and \cite{JaroszVidalKanzieper}.
In addition we have derived for the
quaternion ensemble a different representation
of j.p.d.f. and found a new type of random matrix model.
These results can be used as a starting point for further analyzing 
thermal transport through such quantum dots.

\vspace{.5cm}

{\bf Aknowledgements.} We thank the DFG graduate college IRTG 1132 for suppport.

\newpage
\begin{appendix}
\section{Polar Decomposition}
\label{sec:polardecomposition}
In this section we will discuss some details about the polar decomposition and the uniqueness of the decomposition.
The scattering matrix can belong to $SO(N)$ or $Sp(N)$.
The polar decomposition for the scattering matrix states that it can 
be decomposed as follows.
\begin{align}
 S&=\left(\begin{array}{cc}
    {\bf r}_{n\times n} & {\bf t}_{n\times m} \\
    {\bf t'}_{m\times n}& {\bf r'}_{m\times m} 
   \end{array}\right)\notag\\
   &=\left(\begin{array}{cc}
    U_{1} & 0\\
    0& U_{2} 
   \end{array}\right) 
   \left(\begin{array}{cc}
     r & t \\
    -t'& r' 
   \end{array}\right)
   \left(\begin{array}{cc}
    V_{1}^{\dagger} & 0 \\
    0& V_{2}^{\dagger} 
   \end{array}\right)\notag\\
   &=U 
   \left(\begin{array}{cc}
     r & t \\
    -t'& r' 
   \end{array}\right)
V^{\dagger}\notag
\end{align}
where $r, t, t', r'$ are diagonal matrices and we have taken $n\leq m$.
The matrices $U_{1},V_{1}\in O(n)$ and $U_{2},V_{2}\in O(m)$ when $S\in SO(N)$ while 
the matrices $U_{1},V_{1}\in Sp(n)$, $U_{2},V_{2}\in Sp(m)$ when $S\in Sp(N)$.
Given the unitarity condition on $S$
we have the following relations among the diagonal elements of  $r, t, t', r'$
\begin{align}
\begin{array}{cc}
&r_{j}^{2}+t_{j}^{2}=1  \\
&t'_{j}=t_{j} \\
\text{for $j\leq n$}&r'_{j}=r_{j}\\
\text{for $j> n$} &r'_{j}=1\\
\end{array}
\end{align}
However this decomposition is not unique.
The matrix $r'$ has all diagonal elements equal to $1$ when $j>n$ which means it is invariant
under a unitary transformation in this sector. 
The rectangular matrices $t$ and $t'$ is fulled with $0$'s in this sector.
Thus we can restrict $U_{2}$ to the coset space
$O(m)/O(m-n)$ when the scattering matrix is in $SO(N)$
and to the coset space $ Sp(m)/Sp(m-n)$ when the scattering matrix is in $Sp(N)$.
The number of degrees of freedom of $SO(N)$
is given by $\frac{N(N-1)}{2}$.
In our  parametrization we have 
$\frac{n(n-1)}{2}$ degrees of freedom for $U_{1},V_{1}$,
$\frac{m(m-1)}{2}$ degrees of freedom for $V_{2}$
and $\frac{m(m-1)}{2}-\left(\frac{(m-n)(m-n-1)}{2}\right)$ degrees of freedom for $U_{2}$.
Adding to these the $n$ degrees of freedom from the $r_{j}$ variables we have in total
$\frac{n^{2}+m^{2}-n-m}{2}
+mn
$
which accounts for all the number of degrees of freedom of $SO(N)$, $\frac{N(N-1)}{2}$. 

\noindent A similar situation presents itself for the decomposition of $Sp(N)$.
The number of degrees of freedom for $Sp(N)$ is $N(2N+1)$.
To make the parametrization unique we take $U_{1}\in Sp(n)$,
$V_{1}\in Sp(n)/Sp(1)^{n}$, $U_{2}\in Sp(m)/Sp(m-n)$ and $V_{2}\in Sp(m)$
Summing up the degrees of freedom we have $2n^{2}+2m^{2}
+4mn+m+n$ which correspond to the $N(2N+1)$ degrees of freedom of $Sp(N)$.
A unique parametrization is necessary to compute the Jacobian. However
the scattering matrix is invariant under the subgroup $O(m-n)$ for the case or 
$SO(N)$ and invariant under $Sp(m-n)$ and $Sp(1)^{n}$ in the case of $Sp(N)$.
This means that we can extend the integration over the coset space to the group
, the difference being a proportionality constant.
More precisely we have for every matrix $U$ of the form
\begin{align}
 U&=\left(\begin{array}{cc}
    \mathbb{I}_{m} & 0\\
    0 & U'
   \end{array}
\right)\notag\\
U'&\in O(m-n) , \notag
\end{align}
we have 
\begin{align}
F\left( 
   \left(\begin{array}{cc}
     U_{1}rV_{1} & U_{1}tV_{2} \\
    -U_{2}t'V_{1}& U_{2}r'V_{2} 
   \end{array}\right)
 \right)= F\left( 
   \left(\begin{array}{cc}
     U_{1}rV_{1} & U_{1}tV_{2} \\
    -U_{2}Ut'V_{1}& U_{2}Ur'V_{2} 
   \end{array}\right)
 \right) . \notag
\end{align}
Therefore the identity holds when integrating over $U'$
\begin{align}
F\left( 
   \left(\begin{array}{cc}
     U_{1}rV_{1} & U_{1}tV_{2} \\
    -U_{2}t'V_{1}& U_{2}r'V_{2} 
   \end{array}\right)
 \right)= 
 \int_{O(m-n)} d\mu(U')F\left( 
   \left(\begin{array}{cc}
     U_{1}rV_{1} & U_{1}tV_{2} \\
    -U_{2}Ut'V_{1}& U_{2}Ur'V_{2} 
   \end{array}\right)
 \right) . \notag
\end{align}
We have then
\begin{align}
& \int_{SO(N)} d\mu(S)F(S)\notag\\
 &=
\int dr_{j}J(r_{j}) \int_{O(n)} d\mu(U_{1}) d\mu(V_{1})d\mu(V_{2})
 \int_{O(n)/O(m-n)} d\mu(U_{2})
F\left( 
   \left(\begin{array}{cc}
     U_{1}rV_{1} & U_{1}tV_{2} \\
    -U_{2}t'V_{1}& U_{2}r'V_{2} 
   \end{array}\right)
 \right)\notag\\
 &=
\int dr_{j}J(r_{j}) \int_{O(n)} d\mu(U_{1}) d\mu(V_{1})d\mu(V_{2})
 \int_{O(n)/O(m-n)} d\mu(U_{2})
\int_{O(m-n)} d\mu(U) F\left( 
   \left(\begin{array}{cc}
     U_{1}rV_{1} & U_{1}tV_{2} \\
    -U_{2}Ut'V_{1}& U_{2}Ur'V_{2} 
   \end{array}\right)
 \right) \notag\\
 &= 
\int dr_{j}J(r_{j}) \int_{O(n)} d\mu(U_{1}) d\mu(V_{1})d\mu(V_{2})
 d\mu(U_{2})
F\left( 
   \left(\begin{array}{cc}
     U_{1}rV_{1} & U_{1}tV_{2} \\
    -U_{2}t'V_{1}& U_{2}r'V_{2} 
   \end{array}\right)
 \right) 
\end{align}

\section{Theory of symmetric functions}
\label{sec:TSF}
We use here the theory of symmetric functions
to expand a given symmetric function of multiple  
variables $f(x_{1},\cdots, x_{n})$ in terms of 
Jack polynomials and subsequently integrate 
using some known integration properties of these polynomials.
\subsection{Preliminaries and notation}
A set of non increasing integers 
$\lambda=(\lambda_{1},\lambda_{2},\cdots,\lambda_{l}) 
$ is called a partition of $\kappa$ if 
\begin{align}
\lambda&=(\lambda_{1},\lambda_{2},\cdots, \lambda_{l})\notag\\ 
\lambda_{j}&\geq \lambda_{j+1}\notag\\
\sum_{j=1}^{l}\lambda_{j}&=\kappa. \notag
\end{align}
$\kappa$ 
is called the weight of the partition and the length 
of the partition, $l(\lambda)=l$, is the number of integers $\lambda_{j}$.
Often one also writes  $(\lambda_{1},\lambda_{2},\cdots, \lambda_{l},0,\cdots,0)=
(\lambda_{1},\lambda_{2},\cdots, \lambda_{l})$.
To each partition there is an associated diagram made of boxes.
For a partition $\lambda$ there are $\lambda_{1}$ boxes in the first row,represented by a diagram 
$\lambda_{2}$ boxes in the second row and so on. The length, $l(\lambda)$, then denotes the amount of rows.
Each box is then denoted by two coordinates $s=(i,j)$.
The arm of a box $a_{\lambda}(s)$ is equal to the number of boxes directly to the right 
of the box $s=(i,j)$.
The leg of a box $l_{\lambda}(s)$ is equal to the number of boxes directly below 
of the box $s=(i,j)$.
Similarly the co-arm and co-leg are defined as the boxes directly to the left and above 
the box $s=(i,j)$
The conjugate of a partition $\lambda$ is a partition denoted by $\lambda^{T}$ and defined as
\begin{align}
 \lambda^{T}_{k}=\# \left\{\lambda_{j}\in \lambda : \lambda_{j}\geq k\right\}\notag
\end{align}
The generalized Pochhammer symbol is given by 
\begin{align}
 \left[u\right]_{\lambda}^{(\alpha)}
 =& \prod_{j>1}\frac{\Gamma\left(u-\frac{j-1}{\alpha}+\lambda_{j}\right)}{\Gamma\left(u-\frac{j-1}{\alpha}\right)}
\end{align}
We define the following coefficients
\begin{align}
 d_{\lambda}(\alpha)=&\prod_{s\in \lambda} \left(\alpha a(s)+\alpha +l(s)+1\right)\notag\\
 d'_{\lambda}(\alpha)=&\prod_{s\in \lambda} \left(\alpha a(s)+\alpha +l(s)\right)\notag\\
  e_{\lambda}(\alpha,n)=&\prod_{s\in \lambda} \left(\alpha a'(s)+\alpha+n -l'(s)\right)\notag\\
  e'_{\lambda}(\alpha,n)=&\prod_{s\in \lambda} \left(\alpha a'(s)+\alpha+n -l'(s)-1\right)\notag\\
    h_{\lambda}(\alpha)=&\prod_{s\in \lambda} \left(\alpha a(s)+l(s)+1\right)\notag\\
  b_{\lambda}(\alpha,n)=&\prod_{s\in \lambda} \left(\alpha a'(s)+n -l'(s)\right)\notag
\end{align}
We have then in terms of the Pochhammer symbol 
\begin{align}
 b_{\lambda}(\alpha,n)&=\alpha^{|\lambda|}\left[\frac{n}{\alpha}\right]_{\lambda}^{(\alpha)}\\
 e_{\lambda}(\alpha,n)&=\alpha^{|\lambda|}\left[1+\frac{n}{\alpha}\right]_{\lambda}^{(\alpha)} \\
 e'_{\lambda}(\alpha,n)&=\alpha^{|\lambda|}\left[1+\frac{n-1}{\alpha}\right]_{\lambda}^{(\alpha)}\label{eq:e}
\end{align}
\subsubsection{A combinatorial identity}
We have the following identity for $\alpha=1/2$ 
\begin{align}
S_{\lambda \cup \lambda}(\mathbb{I}_{2n})=
\frac{e'_{\lambda}(1/2,n) b_{\lambda}(1/2,n) }{d'_{\lambda}(1/2) h_{\lambda}(1/2)} 
  \notag
\end{align}
The right hand side can be written down in terms of Pochhammer symbols 
and because we have on the left hand side a schur function of the identity we can express this side 
also in terms of the Pochhammer symbol.
\begin{align}
  \frac{\left[2n\right]^{(1)}_{\lambda \cup \lambda}}{h_{\lambda \cup \lambda}(1)}
=
\frac{\left(1/2\right)^{2|\lambda|}
\left[2n-1\right]_{\lambda}^{(1/2)}
\left[2n\right]^{(1/2)}_{\lambda}
}{d'_{\lambda}(1/2) h_{\lambda}(1/2)} 
  \label{eq:id}
\end{align}
This identity holds for $n$ integer and we will show now it holds for $n$ real. 
We denote by $\mu_{j}$ the $j^{\text{th}}$ integer of the partition 
$\lambda \cup \lambda$. Meaning $\lambda \cup \lambda=\left\{\mu_{1},\mu_{2},\cdots \mu_{l(\lambda\cup \lambda)}\right\}$.
Thus
$\mu_{2j-1}=\lambda_{j}$ and $\mu_{2j}=\lambda_{j}$.
The Pochhammer symbol on the right can be decomposed as follows
\begin{align}
\left[2x\right]^{(1)}_{\lambda \cup \lambda}
&=\prod_{j}^{l(\lambda \cup \lambda)}
\frac{\Gamma\left(2x-j+1+\mu_{j}\right)}{\Gamma\left(2x-j+1\right)} \notag\\
&=\prod_{j}^{l(\lambda )}
\frac{\Gamma\left(2x-(2j-1)+1+\lambda_{j}\right)}{\Gamma\left(2x-(2j-1)+1\right)}
\prod_{j}^{l(\lambda)}
\frac{\Gamma\left(2x-2j+1+\lambda_{j}\right)}{\Gamma\left(2x-2j+1\right)} \notag\\
&=\prod_{j}^{l(\lambda )}
\frac{\Gamma\left(2x-2(j-1)+\lambda_{j}\right)}{\Gamma\left(2x-2(j-1)\right)}
\prod_{j}^{l(\lambda)}
\frac{\Gamma\left(2x-1-2(j-1)+\lambda_{j}\right)}{\Gamma\left(2x-1-2(j-1)\right)} \notag\\
\left[2x\right]^{(1)}_{\lambda \cup \lambda}
&=\left[2x\right]^{(1/2)}_{\lambda}
\left[2x-1\right]^{(1/2)}_{\lambda}
\end{align}
We have not assumed $x$ to be an integer here so this identity is valid for $x$ real.
If we use this in the identity, Eq. (\ref{eq:id}) 
above, we gather
\begin{align}
\frac{\left(1/2\right)^{2|\lambda|}
}{d'_{\lambda}(1/2) h_{\lambda}(1/2)} 
  =
  \frac{1}{h_{\lambda \cup \lambda}(1)}
  \notag
\end{align}
This no longer depends on $n$ and so it is a combinatorial relation.
Thus we have by multiplying the right by $\left[2x\right]^{(1)}_{\lambda \cup \lambda}
$ and the left by $\left[2x\right]^{(1/2)}_{\lambda}
\left[2x-1\right]^{(1/2)}_{\lambda}$
\begin{align}
  \frac{\left[2x\right]^{(1)}_{\lambda \cup \lambda}}{h_{\lambda \cup \lambda}(1)}&=
\frac{\left(1/2\right)^{2|\lambda|}
\left[2x\right]^{(1/2)}_{\lambda}
\left[2x-1\right]^{(1/2)}_{\lambda}
}{d'_{\lambda}(1/2) h_{\lambda}(1/2)} \notag\\
&=
\frac{e'_{\lambda}(1/2,x) b_{\lambda}(1/2,x)
}{d'_{\lambda}(1/2) h_{\lambda}(1/2)} 
\label{eq:identityquaternion}
\end{align}

\subsection{Jack Polynomials}

The Jack  polynomials, denoted by 
$ P^{(\alpha)}_{\lambda}\left(x_{1},\cdots x_{n} \right)$,
are multi-variable polynomials which are symmetric under the permutation 
of the variables and they form a basis for expanding other symmetric functions.
$\alpha$ is a real index and in our case will be related to some kind of symmetry
but we can also view them as a different set of symmetric polynomials that is orthogonal
with respect to a different scalar product. When $\alpha=1$ the Jack polynomial is equal to the Schur
polynomial,
$ P^{(\alpha)}_{\lambda}\left(x_{1},\cdots x_{n} \right)=S_{\lambda}\left(x_{1},\cdots x_{n}\right)$.
The variables of the Jack polynomials can also be seen as the eigenvalues of a
matrix, which is our case.
One has then the notation
\begin{align}
P^{(\alpha)}_{\lambda}\left(X \right)= P^{(\alpha)}_{\lambda}\left(x_{1},\cdots x_{n} \right)\notag
\end{align}
with $x_{j}$ the eigenvalues of the matrix $X$.
For our purposes we are only interested
in the Jack polynomials with $\alpha=2,1,\frac{1}{2}$ which 
corresponds in the random matrix perspective to
$\beta=1,2,4$ respectively ( $\alpha=\frac{2}{\beta}$).
The Jack polynomials evaluated at the identity matrix is known and given in terms of
the Pochhammer symbol as
\begin{align}
P^{(\alpha)}_{\lambda}\left(\mathbb{I}_{n}\right)=&\frac{b_{\lambda}\left(\alpha,n\right)}{h_{\lambda}(\alpha)} 
=\left(\alpha\right)^{|\lambda|}\frac{
\left[\frac{n}{\alpha}\right]_{\lambda}^{(\alpha)}
}{h_{\lambda}(\alpha)}\label{eq:jackidentity}
\end{align}
In addition there exist relations between the different Jack polynomials evaluated at the identity and the Schur
polynomials.
Let us define the following partitions constructed from a partition $\lambda$.
\begin{align}
 \lambda=&(\lambda_{1},\lambda_{2},\cdots) \notag\\
 2\lambda=& (2\lambda_{1},2\lambda_{2},\cdots) \notag\\
 \lambda\cup \lambda=& (\lambda_{1},\lambda_{1},\lambda_{2},\lambda_{2}\cdots) \notag
\end{align}
Then the following identities hold
\begin{align}
\frac{e'_{\lambda}(\alpha,n) }{d'_{\lambda}(\alpha)} P_{\lambda}^{(\alpha)}(\mathbb{I}_{n})
=&\left\{ 
\begin{array}{cc}
  S_{\lambda}(\mathbb{I}_{n}) &  \alpha=1\\
    S_{2\lambda}(\mathbb{I}_{n}) &\alpha=2\\
  S_{\lambda \cup \lambda}(\mathbb{I}_{2n}) &  \alpha=\frac{1}{2}
 \end{array}
 \right.
\label{eq:schurjack}
\end{align}
We are mainly interested in the expansion of  the determinant raised to some power.
In terms Schur polynomials it is given as follows
\begin{align}
 \det\left[1-X\right]^{-a}&=\sum_{\lambda}S_{\lambda}(\mathbb{I}_{a})
 S_{\lambda}(X)\label{eq:detexpansion}\\
 &=\sum_{\lambda}
 \frac{\left[a\right]^{(1)}_{\lambda}}{d'_{\lambda}(1)}
 S_{\lambda}(X)
\end{align}
In this expansion the coefficients in front of the Jack polynomials
are given themselves in terms of Jack polynomials evaluated
at the identity.
The Zonal spherical functions defined by Macdonald \cite{McDonalds}, $\Omega^{(\alpha)}_{\lambda}(x)$,
are defined by the following integration property
\begin{align}
\int_{O(n)}dU\Omega^{(2)}_{\lambda}(AUB)&=\Omega^{(2)}_{\lambda}(A)\Omega^{(2)}_{\lambda}(B)\label{eq:zonalO}\\
\int_{Sp(n)}dU\Omega^{(1/2)}_{\lambda}(AUB)&=\Omega^{(1/2)}_{\lambda}(A)
\Omega^{(1/2)}_{\lambda}(B) \label{eq:zonalS} 
\end{align}
and are give in terms of Jack polynomials as
\begin{align}
 \Omega^{(\alpha)}_{\lambda}(x_{n})&=\frac{P^{(\alpha)}_{\lambda}(x_{n}x^{\dagger}_{n})}{P^{(\alpha)}_{\lambda}(\mathbb{I}_{n})} \label{eq:spherical}
\end{align}
and can be related to the Schur functions via an integration over a group 
that depends on the symmetry index $\beta$.
\begin{align}
 \Omega^{(2)}_{\lambda}(x_{n})
 &=\int_{O(n)}dk S_{2\lambda}(kx)\label{eq:sphericalO} \\
 \Omega^{(1/2)}_{\lambda}(x)& =\int_{Sp(n)} dk S_{\lambda \cup \lambda}(kx)
 \label{eq:sphericalS}
\end{align}
A careful use of the expansion and of the integration theorems
(\ref{eq:zonalO}) , (\ref{eq:zonalS}) , (\ref{eq:sphericalO}) and (\ref{eq:sphericalS}),
is what ultimately will allow us to compute the j.p.d.f.
The Hypergeometric Function of two Matrix Argument, HFMA${}_{2}$,
is given as follows
\begin{align}
 {}_{2}\mathcal{F}_{1}\left(a,b;c\big|X,Y\right)
= \sum_{\lambda}\frac{\alpha^{|\lambda|}}{d'_{\lambda}(\alpha)}
\frac{\left[a\right]_{\lambda}^{(\alpha)}\left[b\right]_{\lambda}^{(\alpha)}}{
\left[c\right]_{\lambda}^{(\alpha)}}
\frac{P_{\lambda}^{(\alpha)}\left(X\right)
P_{\lambda}^{(\alpha)}\left(Y\right)}{P_{\lambda}^{(\alpha)}\left(\mathbb{I}\right)}.
\label{eq:HFMArg2}
\end{align}
The Hypergeometric Function of one Matrix Argument, HFMA${}_{1}$,
is given as follows
\begin{align}
 {}_{2}F_{1}\left(a,b;c\big|X\right)
= \sum_{\lambda}\frac{\alpha^{|\lambda|}}{d'_{\lambda}(\alpha)}
\frac{\left[a\right]_{\lambda}^{(\alpha)}\left[b\right]_{\lambda}^{(\alpha)}}{
\left[c\right]_{\lambda}^{(\alpha)}}
P_{\lambda}^{(\alpha)}\left(X\right) ,\label{eq:HFMArg1}
\end{align}
and $ {}_{2}F_{1}\left(a,b;c\big|X\right)= {}_{2}\mathcal{F}_{1}\left(a,b;c\big|X,\mathbb{I}\right)$.

\section{Average of characteristic polynomials}
We wish to compute the average of characteristic polynomials
of Jacobi Ensembles.
\begin{align}
 \prod_{j=1}^{p} \int_{-1}^{1}
 d\lambda_{j}\left|\Delta\left(\lambda_{j}\right)\right|^{\beta} 
 \left(1-\lambda_{j}\right)^{a} 
\left(1+\lambda_{j}\right)^{b} \prod_{k,j}^{n,p}\left(v_{k}-\lambda_{j} \right)\label{eq:CharacteristicP}
\end{align}
We follow here \cite{ForresterNagao}
where it was found that these averages of characteristic polynomials can be written down as a pfaffian.
We note that in \cite{Borodin} another pfaffian representation was derived.
We introduce the skew-orthogonal polynomials $q_{l}(x)$ which satisfy
the following orthogonality relations 
\begin{align}
\left<q_{2l},q_{2p}\right>&=0\notag \\ 
\left<q_{2l+1},q_{2p+1}\right>&=0\notag \\
\left<q_{2l},q_{2p+1}\right>&=r_{l}\delta_{lp}\notag 
\end{align}
with the scalar product
\begin{align}
 \left<f,g\right>&=\int_{-1}^{1} dvduf(v)f(u)
w(v)w(u) 
\text{sign}(x-y) \\
w(u)&=  \left(1+u\right)^{a} 
\left(1-u\right)^{b} \label{eq:skewscalarweight}
\end{align}
There are four possible cases depending if $p$ and $n$ are even or odd.
In our case $p$ will be even and $n$ arbitrary.
When $p$ is even and $n$ even
 we have 
\begin{align}
\left<\prod_{k=1}^{n}\det\left[v_{k} -Y\right]\right>
=&c_{n,p}
\frac{
\text{Pf}_{j,k\leq n+2}\left[f_{jk}
\right]}{\Delta(v_{k})}\label{eq:pfaffian}
\end{align}
with $f_{jk}$  a $n\times n $ anti symmetric matrix with the following entries
\begin{align}
\begin{array}{cc}
f_{j,k}=F(v_{j},v_{k}) &  
\end{array}\label{eq:fjk}
\end{align}
with 
\begin{align}
F(v,u)
&=
\sum_{l=1}^{\frac{p+n}{2}}\frac{1}{2r_{l-1}}\left(q_{2l-2}(u)q_{2l-1}(v)-q_{2l-2}(v)q_{2l-1}(u)\right)  \label{eq:Feven}
\end{align}
When $p$ is even and $n$ is odd
 we have 
\begin{align}
\left<\prod_{k=1}^{n}\det\left[v_{k} -Y\right]\right>
=&c_{n,p}
\frac{
\text{Pf}_{j,k\leq n+1}\left[f_{jk}
\right]}{\Delta(v_{k})}\label{eq:pfaffian2}
\end{align}
with $f_{jk}$  a $n+1\times n+1$ anti symmetric matrix with the following entries
\begin{align}
\begin{array}{cc}
f_{1,1}=0& \\ 
f_{j,k}=F(v_{j-1},v_{k-1}) &  \text{for $j,k=2,\cdots n+1$}\\
f_{1,j+1}= q_{n+p-1}(v_{j-1}) &\text{for $j=1,\cdots n$}
\end{array}\label{eq:fjk2}
\end{align}
with 
\begin{align}
F(v,u)
&=
\sum_{l=1}^{\frac{p+n-1}{2}}\frac{1}{2r_{l-1}}\left(q_{2l-2}(u)q_{2l-1}(v)-q_{2l-2}(v)q_{2l-1}(u)\right)  \label{eq:Fodd}
\end{align}

\end{appendix}

\end{document}